

 \def\frac#1#2{{\textstyle{#1\over #2}}}
 \def\vphi{\vec\phi}
 
 \def\sprod{\mathinner{\ldotp}}

 \def\modsq#1{{\vert #1 \vert}^2}

\pubnum{CRM-2373}
\titlepage
\title  {Solutions of (2+1)-dimensional spin systems}
\author {Z. Thomova\break
P. Winternitz\break }
\address{Centre de Recherches Math{\'e}matiques\break
Universit\'e de Montr\'eal\break
CP 6128, Succ. Centre Ville\break
Montr\'eal (QC) H3C 3J7, Canada\break
thomovaz@crm.umontreal.ca\break
wintern@crm.umontreal.ca\break}
\author{W.J. Zakrzewski} 
\address{Department of Mathematical Sciences\break
 University of Durham \break
 Durham, DH1 3LE, UK \break
W.J.Zakrzewski@durham.ac.uk\break}

\abstract{We use the methods of group theory to reduce the equations 
of motion of two spin systems in (2+1) dimensions to sets of 
coupled ordinary differential equations. We present solutions 
of some classes of these sets and discuss their physical significance.

Les m\'ethodes de la th\'eorie des groupes sont utilis\'ees pour r\'eduire les 
\'equations du mouvement de deux syst\`emes de spins de dimensions (2+1) \`a des syst\`emes d'\'equations diff\'erentielles ordinaires. Les solutions de certaines classes de ces syst\`emes sont present\'ees et les aspects physics sont discut\'es.}

\chapter {Introduction}

In this paper we look for solutions of the equations of the
Landau-Lifshitz model
(with, perhaps, nonvanishing anisotropy) and of a nonlinear vector
diffusion equation. The equations are given, respectively, by

$${\partial\vphi\over \partial t} = \vphi\times \vec F\eqn\eun$$
and 
$${\partial\vphi\over \partial t} =  \vec F-\vphi(\vphi\sprod \vec F),\eqn\edeux$$
where $\vec F$ is given by
$$
\vec F=\Delta \vec \phi + (A\phi_3 + B)\vec e_3,\eqn\etrois$$
where $\vec e_3$ is a unit vector in the $3\sp{rd}$ direction in the 
$\vec \phi $ space and
$\vphi$ satisfies $\vphi\sprod\vphi=1$. $A$ and $B$ are possible anisotropy 
coefficients. 

The motivation for this work comes from the original observation
made by Landau and Lifshitz\Ref\rlandau{L. Landau, E. Lifshitz:
Physik A (Soviet Union) 8 (1935) 153} in their study of the 
ferromagnetic continuum.
They pointed out that for phenomena
 for which substantial spatial variations
 occur only over a large number of lattice spacings, 
 we can use the continuum approximation. They showed that a
ferromagnetic medium is characterised by the magnetisation vector
$\vec M$ (like the vector $\vphi$ above) which precesses around
the effective magnetic field and so obeys, what is now called the
Landau-Lifshitz equation, namely \eun. Since the original work 
of Landau and Lifshitz many papers have been written on the subject\Ref\rnikos{see {\it e.g.}
N. Papanicolaou
and T.N. Tomaras: Nucl. Phys. B 360 (1991), 425}
and the equation has been modified by the inclusion of various 
additional terms to $\vec F$. It has  been used to describe the  
 dynamics 
of magnetic bubbles in a ferromagnetic continuum
and also of vortices in HeII or in a superconductor\refmark\rnikos.
Various studies of the dynamics of such topological soliton-like structures
have been performed both theoretically and experimentally\Ref\rrev{A.P. Malozemoff, J.C. Slonczewski: Magnetic domain walls in bubble
materials. New York: Academic Press 1979} \Ref\rrevtwo{T.H.~O'Dell: Ferromagnetodynamics,
the dynamics of magnetic bubbles, domains and domain walls. New York: Wiley 1981} and they have exhibited many interesting, and perhaps
unexpected, phenomena - like the skew deflection 
of these structures under the influence
of a magnetic field gradient which resembles the more familar Hall motion
of electrons in external magnetic and electric 
fields\Ref\rhall{V.G.~Bar'yakhtar
et al: Sov. Phys. Usp. 20 (1977) 298 }.

A recent work of Papanicolaou and Tomaras\refmark\rnikos, as well 
as some earlier work of other people\Ref\rHaldane{F.D.M.~Haldane: 
Phys. Rev. Lett. 57 (1986) 1488. F.G. Mertens et al:
 Nonlinear Coherent Structures in Physics and Biology ed. K.H.~Spatschek, F.G.~Mertens  Plenum (1993)} has shown that many
experimentally observed facts can indeed be explained using the Landau-Lifshitz equation. Much of the work involved deriving various
conserved quantities describing these structures
and then using them to restrict the description of the dynamics.
All this work has provided further evidence  as to 
the relevance of the 
Landau-Lifshitz equation to the description of physical
phenomena.
However, as the Landau-Lifshitz equation is 
quite complicated, only some 
results were obtained in an analytical form.
Most more recent studies\Ref\rnikosa{N.~Papanicolaou: Physica D 74 (1994) 107.\hfill \break 
 N.~Papanicolaou, W.J.~Zakrzewski: Physica D 80 (1995) 225 Phys. Lett. A 210 (1996) 328 } involved numerical simulations. 

The vector nonlinear diffusion equation \edeux\ has less obvious
physical applications but it has been used\Ref\rAndrew{see  e.g. A.J. Bray, K. Humayan: J. Phys. A 23 (1990) 5897 \hfill \break M. Zapotocky, W.J. Zakrzewski: Phys.Rev. E 51 (1995) R5189 }
in the study of phase ordering kinetics where one investigates
 the time evolution
of a system quenched from the disordered into an ordered phase.
This topic has
attracted considerable attention in recent years\Ref\rbray{ For a recent review of the theory of phase
ordering, see {\it e.g.} A.J.~Bray
: Advances in Physics (in press)}.
In fact, it has been shown that many features 
of phase ordering in systems supporting
topologically stable defects (for example, in systems described by the
O($N$) vector model in $d$ dimensions with $d \le N$\Ref\rhum{A.J.~Bray, K.~Humayun:
 Phys. Rev. E 47 (1993) 9} \Ref\thumtwo{B.~Yurke et al.: Phys. Rev. E 47 (1993) 1525}), or
in two and three-dimensional nematic liquid crystals\Ref\rmartin{M.~Zapotocky, P. M.~Goldbart, N.~Goldenfeld: Phys. Rev. E 51 (1995) 1216 }
can be understood theoretically by investigating the dynamics of the 
numerous topological defects generated during the quench.
 A special and interesting case is that of the O($3$)
model system in $2$ spatial dimensions. It supports topologically
stable, but non-singular objects which, in the
condensed matter community language, are called topological textures. 
Such
systems were studied numerically in ref. [\rAndrew].

Given the paucity of analytical results
for both equations \eun\ and \edeux\ (especially involving the dynamics)
one of the aims of this paper is to see what time dependent solutions
can be found using 
 the group theoretical
method of symmetry reduction\Ref\rolver{P.J.~Olver: Applications of Lie Groups
 to Differential Equations. New York: Springer 1986} \Ref\rwinternitz{P.~Winternitz: Lie Groups and Solutions of 
Nonlinear Partial Differential Equations; in Integrable Systems,Quantum Groups and Quantum Field Theories; ed. A.~Ibort and M.A.~Rodriguez Dordrecht: Kluwer 
Academic Publihers 1992} \Ref\rgaeta{G.~Gaeta: Nonlinear Symmetries and Nonlinear Equations. Dordrecht: Kluwer Academic Publishers 1994} .
This method exploits the symmetry of the original equations to find solutions
invariant under some subgroup (the classic example one can give here involves 
seeking solutions in three dimensions which are rotationally invariant).
The method puts all such attempts on a unified footing
and it has been applied with success to many equations \Ref\rrogers{C.~Rogers,W.F.~Ames: Nonlinear Boundary Value Problems in Science and Engineering. San Diego: Academic Press 1989} .
 The method gives equations whose
solutions represent specific solutions of the full equations; the solutions
are determined locally and the method does not tell us whether these solutions
are stable or not with respect to any perturbations.

In a recent paper\Ref\rourpaper{A.M.~Grundland, P.~Winternitz, W.J.~Zakrzewski: J. Math. Phys. 37 (1996) 1501}, two of us (PW and WJZ) together with M. Grundland, have applied this technique to looking for solutions
of the relativistic $CP\sp{1}$ model. 

In this paper we investigate solutions of \eun\ and \edeux. We are particularly
interested in time dependent solutions; all time independent
solutions of \eun\ and \edeux\ (when there is no anisotropy) are also the time 
independent solutions of the relativistc model and so can be found
in ref [\rourpaper].

Like in the relativistic $CP\sp{1}$ model studied before,
in order to perform the 
symmetry reductions, 
we have to decide what variables to use. To avoid having to use
the constrained variables ($\vec\phi$) it is convenient to use 
the $W$ formulation 
of the model which involves the stereographic projection of the sphere 
$\vec{\phi}\cdot\vec{\phi}=1$ onto the complex plane.
In this formulation
 instead of using the $\vec\phi$ fields, we express
 all the dependence on $\vec\phi$ in terms of 
their stereographic projection onto
 the complex plane $W$. The $\vec\phi$ fields are then related to $W$ by
$$
 \phi_1\,=\,  {W + W^* \over 1 + \modsq{W}},\,\,\,\,\,
 \phi_2\,=\,i {W - W^* \over 1 + \modsq{W}},\,\,\,\,\,
 \phi_3\,=\,  {1 - \modsq{W} \over 1 + \modsq{W}}.
\eqn\eWToPhi
$$

To perform our analysis it is convenient to use the
polar version of the $W$ variables;
\ie\ to put $W=R\,\exp{iQ}$ and then study the equations
for $R$ and $Q$. The advantage of this approach is that the equations
 become simple; the disadvantage comes from having to pay attention
that $R$ is real  and $Q$ should be periodic with
a period of $2\pi$. (If the period is not 
$2\pi$ then the solution may become multi-valued) 
Thus if we find solutions
that do not obey these restrictions, then these solutions, however
interesting they may be, cannot be treated as solutions of the original
model.

In the case of the Landau-Lifshitz equation the
 equations for $R$ and $Q$ take the form
$$\partial_{t}R-
 2{(1-R\sp2)\over (1+R\sp2)}\Bigl( 
\partial_x Q \partial_x R +
\partial_y Q \partial_y R\Bigr)-R 
(\partial_{xx}Q+\partial_{yy}Q)  =0\eqn\eradiallandau$$
and
$$\eqalign{\partial_tQ\,=B+A{1-R\sp2\over 1+R\sp2}-&{\partial_{xx}R+\partial_{yy}R\over R}\cr
 +{(1-R\sp2)\over (1+R\sp2)}\Bigl(
(\partial_x Q)\sp2+
(\partial_y Q)\sp2\Bigr)
 +&{2\over (1+R\sp2)}\Bigl( 
(\partial_x R)\sp2
+(\partial_y R)\sp2\Bigr),\cr}\eqn\ephaselandau$$

while for the diffusion case they are
respectively
$$\partial_{t}Q
 -2{(1-R\sp2)\over R(1+R\sp2)}\Bigl( 
\partial_x Q \partial_x R +
\partial_y Q \partial_y R\Bigr)-(\partial_{xx}Q+\partial_{yy}Q)  =0\eqn\ephasediffusion$$
and
$$\eqalign{\partial_tR\,+BR+AR{1-R\sp2\over 1+R\sp2}-&\partial_{xx}R-\partial_{yy}R\cr
 +{(1-R\sp2)R\over (1+R\sp2)}\Bigl(
(\partial_x Q)\sp2+
(\partial_y Q )\sp2\Bigr)
 +&{2R\over (1+R\sp2)}\Bigl( 
(\partial_x R)\sp2
+(\partial_y R)\sp2\Bigr)\,=\,0.\cr}\eqn\eradialdiffusion$$

Note, that, in the Landau-Lifshitz case,
 if we put $R=1$ the equations become $\Delta Q=0$ and $\partial_tQ=B$
which have a very simple solution, and in the diffusion case, we have to set
$B=0$ and then we end up with
$\partial_tQ-\Delta Q=0$ as the equation for $Q$. The latter case
is the nonrelativistic analogue of what was found in the
relativistic case where $R=1$ reduced the equation for $Q$ to the linear wave equation for the phase $Q$.

In the next section we determine the symmetry group
of our equations \ephaselandau\ , \eradiallandau\ and of
\ephasediffusion\ and \eradialdiffusion\ . In the following sections 
we solve the derived equations and discuss their 
solutions.


\chapter{The Symmetry Group and its Two Dimensional Subgroups}

The symmetry group of our systems of equations, respectively
\ephaselandau\ and \eradiallandau\ and
\ephasediffusion\ and \eradialdiffusion\ , can be calculated using
the standard methods \refmark\rolver\ \refmark\rwinternitz\ \refmark\rgaeta\ \refmark\rrogers\ .
We actually made use of a 
 MACSYMA package\Ref\rmaxyma{B.~Champagne, W.~Hereman, P.~Winternitz: Comp. Phys. Commun. 66 (1991) 319 } that provides a simplified and partially
solved set of determining equations. 

Solving the determining equations we find that three different cases must be
distinguished:
\item{1.} $A=B=0$, {\it i.e.} the anisotropy is absent. The Landau-Lifshitz
equation and the diffusion equation have isomorphic symmetry groups,
consisting of translations in space and time directions, rotations in the
$x,y$ plane, dilations and a group of $O(3)$ rotations between the components
of the field $\vphi$. The corresponding Lie algebra $L_1$ has the structure 
of a direct sum
$$ L_1=s(2,1) \quad \oplus \quad O(3).\eqn\esymm$$

Bases for these two algebras are given by the following vector fields, acting
on space-time and on the fields in the $\{R,Q\}$ realisation of eq. \eradiallandau-\eradialdiffusion:
$$ \eqalign{s(2,1):\qquad&P_0=\partial_t,\quad P_1=\partial_x,\quad
P_2=\partial_y,\cr
& L=-x\partial_y+y\partial_x,\quad D=2t\partial_t+x\partial_x+y\partial_y.\cr}
\eqn\es$$

$$\eqalign{su(2),\qquad &X={1\over 2}\Bigl(sinQ \bigl(R-{1\over R}\bigr)\partial_{Q}
+cosQ\bigl(R\sp2+1\bigr)\partial_R\Bigr),\cr
&Y={1\over 2}\Bigl(cosQ \bigl(R-{1\over R}\bigr)\partial_{Q}
-sinQ\bigl(R\sp2+1\bigr)\partial_R\Bigr),\cr
&Z=\partial_{Q}.\cr}\eqn\estwo$$

\item{2.} $A\ne0$.

The symmetry algebra for both equations is reduced to
$$L_2=\{P_0,P_1,P_2,L\}\quad \oplus \quad \{Z\},\eqn\ereduced$$
{\it i.e.} the dilations are absent and the only $\vphi$ rotations
left are those around the third axis ({\it i.e.} around $\phi_3$).

\item{3.} $A=0,\quad B\ne 0$.

The symmetry algebra for the dissipative equations \eradialdiffusion\ and
\ephasediffusion\ is still $L_2$, as in \ereduced. That of the 
Landau-Lifshitz equation is
$$L_3=\{P_0,P_1,P_2,L,\tilde D\}\quad \oplus \quad \{X,Y,Z\},\eqn\ereduceda$$
with 
$$\tilde D=2t\partial_t+x\partial_x+y\partial_y+2Bt\partial_Q.\eqn\ereducedb$$

In order to perform symmetry reduction we need to classify the subalgebras of the symmetry
algebras $L_1,L_2$ and $L_3$. We wish to reduce equations \eradiallandau-\eradialdiffusion\ to ordinary differential equations.
To do this, we will require that the solutions are invariant under a two-dimensional subgroup of the symmetry group. In order to do this
systematically we need to derive a classification of the two dimensional
subalgebras of the symmetry algebra. Moreover, we can restrict ourselves
to subalgebras, all elements of which act nontrivially on space-time,
{\it i.e.} which do not contain any rotations in $\vphi$ space.

The subalgebra classification can be done in an algorithmic way \refmark\rwinternitz; 
the results are quite simple and we present them
without a proof.

\item{1.} $A=B=0$.
Every two-dimensional subalgebra of $L_1$, each element of which acts nontrivially on space-time, is conjugate under the action of the group
of inner automorphisms to one of the following ones
$$\eqalign{A_{2,1}=&\{P_1+aZ,P_2+bZ\},\cr
A_{2,2}=&\{L+aZ,P_0+bZ\},\cr
A_{2,3}=&\{P_0+aZ,P_2+bZ\},\cr
A_{2,4}=&\{P_0-vP_1+aZ,P_2+bZ\},\quad v\ne0,\cr
A_{2,5}=&\{D+bL+aZ,P_0\},\quad b\ne0, \cr
A_{2,6}=&\{D+aZ,P_0\},\cr
A_{2,7}=&\{D+aZ,L+bZ\},\cr
A_{2,8}=&\{D+aZ,P_2\}.\cr}\eqn\eeverything$$

The parameters $a,b$ and $v$ are arbitrary real numbers. In some cases their
ranges can be further constrained but that is not important for our 
purposes.

\item{2.} $A\ne0$.

Every two-dimensional subalgebra of the considered type is conjugate  
to one listed above as $A_{2,1},...A_{2,4}$.

\item{3.} $A=0,\quad B\ne0$.

For the dissipative equations \eradialdiffusion\ and \ephasediffusion\ the subalgebra classes are represented by $A_{2,1},...A_{2,4}$. For the 
Landau-Lifshitz equations \eradiallandau\ and \ephaselandau\ they
are represented by $A_{2,1},...A_{2,8}$ with $D$ replaced by $\tilde D$ and
$P_0$ replaced by $P_0=P_0+bZ$.

We can now proceed to perform various reductions. We are particularly 
interested in reductions that do not result in time independence 
as these were already studied in ref [\rourpaper].


\chapter{Solutions of the Landau-Lifshitz Equation}

\section{General Procedure}
Our aim is to solve the Landau-Lifshitz equations \eradiallandau\ and 
\ephaselandau\ , using the method of symmetry reduction. This involves assuming
 that a solution is invariant under a subgroup $G_{0}$ of the symmetry group $G$, namely one of the two dimensional groups corresponding to the algebras $A_{2,1}, \ldots ,A_{2,8}$ of \eeverything. 
The assumption makes it possible to reduce the partial differential equations \eradiallandau\ and
 \ephaselandau\ to a pair of coupled ordinary differential equations. Whenever possible, we 
 decouple them and find explicit solutions for the functions $R$ and $Q$, hence for $W$, and 
 finally for the vector $\vphi$ figuring in \eun\ .

For all 8 algebras in  \eeverything\ the invariant solution will have the form $$
R(x,y,t)=R(\xi), \qquad Q(x,y,t)=\alpha(\xi)+\beta(x,y,t) \eqn\eRandQ$$
where $\xi$ and $\beta$ are explicitly given and $R(\xi)$ and $\alpha(\xi)$ satisfied coupled ordinary differential equations obtained by substituing \eRandQ\ into \eradiallandau\ and \ephaselandau\ .

The reduced equation \eradiallandau\ is
$$
(\nabla\xi)^2 \alpha_{\xi\xi}+\bigl[2{(1-R^2) \over R(1+R^2)} (\nabla\xi)^2
R_{\xi}+\Delta\xi\bigr]\alpha_{\xi}=
{R_{\xi} \over R} \xi_{t} -2 {(1-R^2) \over R(1+R^2)} R_{\xi} (\nabla\xi,\nabla\beta) - \Delta\beta. \eqn\ereducedlandau$$
For algebra $A_{2,1}$ we have 
$$
\nabla\xi^2=\Delta\xi=\Delta\beta=0, \qquad \xi_{t}=1 
\eqn\ejedna$$
and so \ereducedlandau\ reduces to $R_{\xi}=0$.

In all other cases we have $(\nabla\xi)^2 \neq 0$. Eq. \ereducedlandau\ is a first order linear inhomogeneous equation for $\alpha_{\xi}$. We can integrate it explicitly and obtain $\alpha_{\xi}$ in terms of $R$, whenever the functions $\xi$ and $\beta$ satisfy
$${d \over d\xi}{\left({h \over (\nabla\xi)^2}\right)}=0, \qquad
{d \over d\xi}\left[h{(\nabla\xi,\nabla\beta) \over (\nabla\xi)^2}\right]-{h \Delta\beta
\over (\nabla\xi)^{2}} = 0 \eqn\ebetaxi$$
where 
$$\eqalign{
h(\xi)=1 \quad &\hbox{for}  \quad \Delta\xi=0, \cr
{h'(\xi) \over h(\xi) } ={ \Delta\xi \over (\nabla\xi)^2 } \quad  &\hbox{for} \quad \Delta\xi \neq 0.} \eqn\ehconditions$$

Conditions \ebetaxi\ are always satisfied for the algebras $A_{2,2} \ldots A_{2,6}$, not however for $A_{2,7}$ and $A_{2,8}$. When conditions \ebetaxi\ are satisfied, we can integrate eq. \ereducedlandau\
 once to obtain
$$\alpha_{\xi}={S \over h} {(1+R^2)^2 \over R^2} + \mu {1+R^2 \over R^2} + \nu \eqn\emini$$
where  $S$ is an arbitrary real integration constant and where
 we have 
$$\matrix
{&\mu=  -{\nu \over 2}, \quad  &\nu = 0, \quad &\hbox{for}& A_{2,4} \cr
&\mu=0,\quad &\nu= -{a \over b^2+1}, \quad &\hbox{for}& A_{2,5} \cr
&\mu=0, \quad &\nu={ a\xi \over 1+\xi^{2}}, \quad &\hbox{for}&A_{2,6} \cr
&\mu=0, \quad &\nu=0, \quad &\hbox{for}& A_{2,2}, A_{2,3}. \cr}
\eqn\ecasesmunu$$

Equation \ephaselandau\ for algebras $A_{2,2}, \ldots , A_{2,8}$ is reduced to a second order differential equation for $R(\xi)$, that also involves $\alpha_{\xi}(\xi)$. For reductions corresponding to  Lie algebras $A_{2,2}, \ldots , A_{2,6}$ we can substitute $\alpha_{\xi}$ from 
 \emini\ , to obtain an ordinary differential equation for $R(\xi)$ alone. To transform this equation to a standard form we put
$$R(\xi)= \sqrt{-U(\eta)}, \quad \eta=\int h^{-1}(\xi)d\xi. \eqn\ertou$$

The equation for $U(\eta)$ is then written as
$$U_{\eta\eta}=\bigl({1 \over 2U} + {1 \over U-1} \bigr) U_{\eta}^2 -{2S^2 \over U}
(U+1)(U-1)^3+p{U(U+1) \over U-1} +qU+m(U-1)^2. \eqn\eudotdot$$
Equation \eudotdot\ can be integrated in terms of elliptic functions if $p,q$ and $m$ are constants. This is always the case for algebras $A_{2,3} , \ldots ,A_{2,6}$. In the case of algebra $A_{2,2}$ this is true if we set $A=0, B=b$.

Eq. \eudotdot\ has a first integral that we can write as
$$U_{\eta}^2=-4S^2U^4+K_{1} U^3+KU^2+K_{2} U+K_{3} \eqn\eudotpolynomial$$
where $K$ is an integration constant, and the constants $K_{1},K_{2}$ and $K_{3}$ are related to the coefficients $S,p,q$ and $m$ in \eudotdot\ .

In this article we restrict ourselves to solutions of the Landau-Lifshitz equation that are obtained by solving  \eudotpolynomial\ .

We shall first discuss solutions of \eudotpolynomial\ in general, then run through algebras $A_{2,2}, \ldots , A_{2,6}$ and specify the values of the coefficients in \eudotpolynomial\ in each case, as well as the independent variable $\eta$.

Algebra $A_{2,1}$ leading to a first order equation, will be treated separately.


\section{Solutions of the elliptic function equation}

We shall call \eudotpolynomial\ the ``elliptic function equation".
 Its solutions are of course well known \Ref\rbyrd{P.F.~Byrd, M.D.~Friedman: Handbook of Elliptic Integrals for Engineers and Scientists. Berlin: Springer 1971}. We shall however list those that are relevant in the context of solving 
 \eudotdot\ , and more importantly, the Landau-Lifshitz equation.

Several comments are in order here:
\item{1.} The functions $R(\eta)$ must be real (and nonnegative), hence $U(\eta)$ must be real and nonnpositive.
\item{2.} For $S \neq 0$ the coefficient of the highest power of $U$ in 
\eudotpolynomial\ is nonnegative. this means that all real solutions of  \eudotpolynomial\ are nonsingular.
\item{3.} For $S=0$, $K_{1} \neq 0$ in 
 \eudotpolynomial\ the real solutions of
 \eudotpolynomial\ can be singular. Since we are really interested in the fields $\phi_{i}$ we note that singular solutions of $U$ will give regular functions $\phi_{i}$.
\item{4.} In general, equation \eudotpolynomial\ is solved in terms of Jacobi elliptic functions. However, these reduce to elementary functions whenever the polynomial on the right hand side has multiple roots, or when $S=K_{1}=0$.

Let us run through individual cases.
\item{I.}$S\neq 0$ \hfil\break
We rewrite  \eudotpolynomial\ as 
$$
U_{\eta}^2=-4S^2(U-U_{1})(U-U_{2})(U-U_{3})(U-U_{4}) \eqn\eumultipl$$
\itemitem{1.} $U_{1} \leq U \leq U_{2}=U_{3}=U_{4} < 0$
$$U(\eta)=U_{2}-{U_{2}-U_{1} \over 1+S^2(U_{2}-U_{1})^2(\eta -\eta_{0})^2}\eqn\eblaone$$
this is an algebraic solitary wave, equal to $U_{2}$ for $\eta \rightarrow \pm \infty$, and dipping down to $U_{1}$ for $\eta=\eta_{0}$. 

\itemitem{2.} $U_{1}=U_{2}=U_{3} < U \leq U_{4} \leq 0$
$$U(\eta)= U_{1}+{U_{4}-U_{1} \over 1+S^2(U_{4}-U_{1})^2(\eta-\eta_{0})^2}\eqn\eblatwo$$
Also an algebraic solitary wave, rising to $U=U_{4}$ for $\eta=\eta_{0}$, equal to $U_{1}$ for $\eta \rightarrow \pm \infty$.

\itemitem{3.} $U_{1}\leq U < U_{2}=U_{3} <  U_{4},\, \, \, U_{2} \leq 0$
$$\eqalign
{U(\eta)=& U_{2}-{(U_{4}-U_{2})(U_{2}-U_{1}) \over (U_{4}-U_{1})\cosh^2\mu(\eta-\eta_{0})-(U_{2}-U_{1})} \cr
\mu=& S \sqrt{(U_{4}-U_{2})(U_{2}-U_{1})}}\eqn\eblathree$$

\itemitem{4.}$U_{1}<U_{2}=U_{3} < U \leq U_{4} \leq 0$
$$U(\eta)= U_{3}+{(U_{3}-U_{1})(U_{4}-U_{3}) \over (U_{4}-U_{1})\cosh^2\mu(\eta-\eta_{0})-(U_{4}-U_{3})} \eqn\eblafour$$
with $\mu$ as in \eblathree\ .

The last two solutions are solitons, the first one a well, the second a bump.

\itemitem{5.} $U_{1}\leq U \leq U_{2} < U_{3} = U_{4}, \, \, U_{2}\leq 0$
$$\eqalign
{U(\eta)=& U_{4}-{(U_{4}-U_{2})(U_{4}-U_{1}) \over (U_{2}-U_{1})\sin^2\mu(\eta-\eta_{0})+U_{4}-U_{2}} \cr
\mu=& S \sqrt{(U_{4}-U_{2})(U_{4}-U_{1})} \cr}\eqn\eblafive$$

\itemitem{6.} $U_{1}= U_{2} < U_{3} \leq U \leq U_{4} \leq 0$
$$\eqalign
{U(\eta)=& U_{1}+{(U_{4}-U_{1})(U_{3}-U_{1}) \over (U_{4}-U_{3})\sin^2\mu(\eta-\eta_{0})+U_{3}-U_{1}} \cr
\mu=& S \sqrt{(U_{4}-U_{1})(U_{3}-U_{1})} \cr}\eqn\eblasix$$

\itemitem{7.} $U_{1}\leq U \leq U_{2} < U_{3} < U_{4}, \, \, U_{2} \leq 0$
$$\eqalign
{U(\eta)=& U_{4}-{(U_{4}-U_{2})(U_{4}-U_{1}) \over (U_{2}-U_{1})sn^2(\mu(\eta-\eta_{0)},k)+U_{4}-U_{2}} \cr
\mu=& S \sqrt{(U_{4}-U_{2})(U_{3}-U_{1})}, \quad 
 k^2={(U_{4}-U_{3})(U_{2}-U_{1}) \over (U_{4}-U_{2})(U_{3}-U_{1})} \cr} \eqn\eblaseven$$

\itemitem{8.} $U_{1}< U_{2} < U_{3} \leq U \leq U_{4} < 0$
$$
U(\eta)=U_{1}+{(U_{4}-U_{1})(U_{3}-U_{1}) \over (U_{4}-U_{3})sn^2\mu(\eta-\eta_{0})+U_{3}-U_{1}} \eqn\eblaeight$$
with $k^2$ and $\mu$ as in  \eblaseven\ .

\itemitem{9.} $U_{1}\leq U \leq U_{2} \leq 0, \, \, \, U_{3,4}=p\pm iq, \, q>0$
$$\eqalign
{U(\eta)=& {(MU_{1}-NU_{2})cn(\mu(\eta-\eta_{0}),k)+MU_{1}+NU_{2} \over (M-N)cn(\mu(\eta-\eta_{0}),k) +M+N } \cr
M^2=&(U_{2}-p)^2+q^2, \, \, \, \, N^2=(U_{1}-p)^2+q^2 \cr
k^2=&{(U_{2}-U_{1})^2-(M-N)^2 \over 4MN}, \, \, \, \mu=2S \sqrt{MN} \cr}
\eqn\eblanine$$

Solutions \eblafive\ , $\ldots$ , \eblanine\ are periodic. All the elementary solutions can be viewed as limits of solutions \eblaseven\ , \eblaeight\ and \eblanine\ .

\item{II.} $S=0, \, \, \, K_{1} \neq 0$ \hfil\break
Set $$\mu={1 \over 2} \sqrt{|K_{1}|(U_{3}-U_{1}} \eqn\eblabla$$ 
\itemitem{1.} $K_{1} < 0, \, U_{1}=U_{2} < U \leq U_{3} \leq 0$
$$U=U_{3}-(U_{3}-U_{2})\tanh^2\mu(\eta-\eta_{0}) \eqn\eblablaone$$

\itemitem{2.} $K_{1} < 0 ,U < U_{1}=U_{2} <U_{3}, U_{1} \leq 0$
$$U=U_{3}-{(U_{3}-U_{1}) \over \tanh^2\mu(\eta-\eta_{0})} \eqn\eblablatwo$$

\itemitem{3.} $K_{1} < 0 ,U < U_{1}=U_{2} =U_{3}\leq 0$
$$U=U_{1}-\sqrt{{2 \over -K_{1}}}{1 \over (\eta-\eta_{0})^2} \eqn\eblablathree$$

\itemitem{4.} $K_{1} < 0 , U \leq U_{1} < U_{2} = U_{3}, U_{1} \leq 0$
$$U=U_{3}-{U_{3}-U_{1} \over \sin^2\mu(\eta-\eta_{0})} \eqn\eblablafour$$

\itemitem{5.} $K_{1} < 0 , U_{1} < U_{2}< U \leq  U_{3} \leq 0$
$$U=U_{3}-(U_{3}-U_{2})sn^2(\mu(\eta-\eta_{0}),k), \qquad k^2={U_{3}-U_{2} \over U_{3}-U_{1}}  \eqn\eblablafive$$

\itemitem{6.} $K_{1} < 0 , U \leq U_{1} < U_{2} < U_{3}, U_{1} \leq 0$
$$U=U_{3}-{U_{3}-U_{1} \over sn^2(\mu(\eta-\eta_{0}),k)} \eqn\eblablasix$$
$k$ as in \eblablafive\

\itemitem{7.} $K_{1} > 0, U_{1} <U < U_{2} =U_{3} =0$
$$U=U_{1}{1 \over \cosh^2\mu(\eta-\eta_{0})} \eqn\eblablaseven$$

\itemitem{8.} $K_{1} > 0, U_{1} <U < U_{2}<0 <U_{3} $
$$U=(U_{2}-U_{1})sn^2(\mu(\eta-\eta_{0}),k), \qquad k^2={U_{2}-U_{1} \over U_{3}-U_{1}} \eqn\eblablaeight$$

\itemitem{9.} $K_{1}<0, U_{1} \leq 0, U_{2,3}=p \pm iq, q> 0$
$$\eqalign
{U=&U_{1}+A-{2A \over 1-cn(\mu(\eta-\eta_{0}),k)} \cr
A^2=&(p-U_{1})^2+q^2,\quad k^2={A-p+U_{1} \over 2A},\quad \mu=\sqrt{|K_{1}|A} \cr}
\eqn\eblablanine$$

\item{III.} $S=0, K_{1}=0, K \neq 0$
\itemitem{1.} $K >0, U \leq U_{1} < 0 < U_{2}$
$$U=U_{1}- (U_{2}-U_{1})\sinh^2{\sqrt{K} \over 2}(\eta - \eta_{0}) \eqn\ebbbone$$

\itemitem{2.} $K >0, U < U_{1} = U_{2}=0$
$$U=-\exp(-\sqrt{K}(\eta -\eta_{0})) \eqn\ebbbtwo$$

\itemitem{3.} $K <0, U_{1} < U < U_{2} \leq 0$
$$U=U_{1}+ (U_{2}-U_{1})\cos^2{\sqrt{-K} \over 2}(\eta - \eta_{0}) \eqn\ebbbthree$$

\item{IV.} $S=K_{1}=K=0, K_{2} \neq 0$
$$U=-{K_{3} \over K_{2}} +{K_{2} \over 4}(\eta - \eta_{0})^2,\quad K_{2} < 0, \quad K_{3} < 0 \eqn\ebbbb$$

\item{V.} $S=K_{1}=K=K_{2}=0$
$$U=\sqrt{K_{3}} (\eta - \eta_{0}), \quad K_{3} >0 \eqn\ebbbbb$$


\section{Individual reductions}

1. Algebra $A_{2,1}$.

This is an exceptional case when 
\ereducedlandau\ implies $R_{\xi}=0$. We find that the only solution 
for $W$ of \eWToPhi\ is
$$W=R_{0}e^{iQ},\quad Q=ax+by+\bigl(B+{1-R_{0}^2 \over 1+R_{0}^2} (a^2+b^2+A)\bigr) t +\alpha_{0}
\eqn\eatwoone$$
where $R_{0}$ and $\alpha_{0}$ are integration constants.

2. Algebra $A_{2,2}$.

We find 
$$W=R(\rho)\exp{i[\alpha(\rho)+a\phi+bt]}, \qquad \xi=\rho \eqn\eatwotwo$$
where $\rho$ and $\phi$ are polar coordinates. 
The singlevaluedness of $W$ requires
$a$ to be an integer. The phase $\alpha(\rho)$ and variable $\eta$ satisfy 
$$\alpha_{\rho}(\rho)=S{(1+R^2)^2 \over \rho R^2}, \quad \eta=\ln\rho \eqn\ealphadot$$ (see \emini\ ).
 For the function $U(\eta)$ of
 \ertou\ we obtain the elliptic function equation if and only if we set
$$A=0, \qquad b=B \eqn\eabB$$
($A$ and $B$ are defined in \ephaselandau\ ).

We have
$$K_{1}=K_{2}=2a^2+4S^2-{K \over 2}, \quad K_{3}=-4S^2 \eqn\ethirtynine$$
in \eudotpolynomial\ .

For $S \neq 0$ eq. \ethirtynine\ implies that we can have two negative and two positive roots in eq. \eumultipl\ or two negative roots and two complex conjugate ones. These cases lead to real solutions, namely \eblafive\ , \eblaseven\ and \eblanine\ . Note that all of them are periodic.
In particular, for $a=0$ eq. \eumultipl\ always has a double root $U_{3}=U_{4}=1$ and reduces to 
$$U_{\eta}^{2}= -4S^{2}[U^{2}+(1+{K \over 8S^{2}})U+1] \eqn\erootone$$

For $S=0$, $K \neq 4a^2$ we obtain the equation
$$\eqalign
{U_{\eta}^2&=2(a^2-{K \over 4}) U (U-U_{1})(U-U_{2}), \cr
U_{1}U_{2}&=1, \quad U_{1}+U_{2}={2K \over K-4a^2}. \cr} \eqn\eszero$$

The relevant solutions of  \eszero\ in this case are:
\item{1.} $2a^2 \leq K <4a^2, U_{1}<U_{2}<0$ \hfil\break
solution \eblablaeight\ (with $U_{3}=0$). For $K=2a^2$ we have $U=-1$.

\item{2.} $K > 4a^2, U_{1} \leq U_{2} <0$ \hfil\break
Solutions \eblablafive\ , \eblablasix\ , \eblablaone\ and \eblablatwo\ (all with $U_{3}=0$).

\item{3.} $K>4a^2, 0<U_{1}<U_{2}$ \hfil\break
Solutions \eblablafive\ (with $U_{3} \rightarrow U_{2}$,  $U_{2}\rightarrow U_{1}, U_{1}=0$ ).

\item{4.} $K>4a^2, U_{1,2}=p \pm iq, q>0$
solution \eblablanine\ .

For $S=0, K=4a^2$  \eudotpolynomial\ reduces to an elementary one and its solution is
$$R(\rho)=-R_{0}^{2}\rho^{\pm2a} \eqn\efourtyone$$
where $R_{0}$ is an integration constant.

3. Algebra $A_{2,3}$ and $A_{2,4}$

The reduction formulas in both of these cases are
$$W=R(\xi)\exp{i[\alpha(\xi)-at-by]}, \quad \xi=x+vt, \quad \eta=\xi \eqn\ewtwofour$$
with $v=0$ and $v \neq 0 $ for the algebras $A_{2,3}$ and $A_{2,4}$ respectively. Since the Landau-Lifshitz equation is not Galilei invariant, we cannot change the value of $v$ by a group transformation. The transformation \ertou\ leads to \eudotpolynomial\ with
$$\eqalign
{K_{1}&= -{K \over 2} +4S^2 +2(A+b^2) -{v^2 \over 2} +2vS +2(-B+a) \cr
K_{2}&= -{K \over 2} +4S^2 +2(A+b^2) +{3v^2 \over 2} -6vS +2(B-a) \cr
K_{3}&= -(2S-v)^2. \cr} \eqn\ealfabetagama$$
Eq. \emini\  in this case gives
$$\alpha_{\xi}=S{(1+R^2)^2 \over R^2} \eqn\ealphadotfour$$

For $S \neq 0$ we obtain  \eumultipl\ with the constraint
$$U_{1}U_{2}U_{3}U_{4}=\bigl(1-{v \over 2S} \bigr)^2 \eqn\euuuu$$
imposed on these roots. Hence, only even number of roots can be negative (0, 2 or 4). This however means that all solutions  \eblaone\  $\ldots$ \eblanine\ can occur, though in some cases we must impose $U_{4} \leq 0$ ($U_{4}=0$ is allowed for $v=2S$).

For $S=0$, all solutions \eblablaone\ , $\ldots$, \eblablanine\ can occur.

4. Algebra $A_{2,5}$

The reduction formula is
$$W=R(\xi)\exp{i[\alpha(\xi)+{a \over b}\phi+Bt]}, \quad \xi=\ln\rho+{1 \over b}\phi \eqn\ewtwofive$$
and we must set $A=0$ in the Landau-Lifshitz equation. The phase $\alpha(\xi)$ satisfies
$$\alpha_{\xi}=S{(1+R^2)^2 \over R^2}-{a \over b^2+1} \eqn\ealphadotfive$$
(see  \emini\ ).

The function $U=-R^2(\xi)$ satisfies  \eudotpolynomial\ and for $S \neq 0$ we have
$$K_{1}=K_{2}=-{K \over 2} +4S^2 -{2a^2 \over (b^2+1)^2}, \quad K_{3}=-4S^2.
 \eqn\eabgfive$$

The solutions that can occur in this case are \eblathree\ , $\ldots$, \eblanine.
However, there are constraints between various parameters
of the solution which follow from the requirement of singlevaluedness
of $W$.

For $a=0$ eq. \eumultipl\ again reduces to \erootone\ and we only obtain the elementary periodic solutions \eblafive\ .

For $S=0$, $K_{1} \neq 0$ we obtain the equation
$$\eqalign
{U_{\xi}^2=& K_{1} U(U-U_{1})(U-U_{2}), \cr
U_{1}U_{2}=1&, \qquad U_{1}+U_{2}=2\bigl[1+{2a^2 \over K_{1}(b^2+1)} \bigr].
\cr} \eqn\eupolynforszero$$
For $-{a^2 \over b^2+1} < K_{1} < 0$ we have $U_{1} \leq U_{2} < 0$ and solutions \eblablaone\ , \eblablatwo\ , \eblablafive\  and \eblablasix\ are obtained.

For $K_{1} > 0$ we have $0 < U_{1} \leq U_{2}$ but we obtain no real solutions.

For $K_{1} < -a^2/(b^2+1)$ we obtain solutions \eblablanine\ .

Finally, for $S=0$, $K_{1}=0$ the solution is $U=-R_{0}^2 \exp{(\pm \sqrt{K}\xi)}$ and hence
$$R=R_{0}\exp{[\pm {1 \over 2}\sqrt{K}(\ln\rho +{1 \over b} \phi)]} \eqn\eRalphaSzero$$

For $S=0$, $K_{1} \neq0$ equation \eudotpolynomial\ reduces to
$$U_{\xi}^2= K_{1} U(U^2+{K \over K_{1}}U +1), \quad K_{1}=-{K \over 2}-{2a^{2} \over (b^{2}+1)^{2}} \eqn\esomething$$
Real solutions are obtained only for $K_{1} < 0$. More specifically, solutions \eblablasix\
and \eblablanine\ can occur for any $K_{1} < 0$. Solution \eblablafive\ for $K_{1}$
in the range $-{2a^{2} \over (b^{2}+1)^{2}} \leq K_{1} < 0$, \eblablaone\ for
$K_{1}=-{a^{2} \over (b^{2}+1)^{2}}$ and \eblablatwo\ either for $a=0$, 
or $K_{1}=-{a^{2} \over (b^{2}+1)^{2}}$.

5. Algebra $A_{2,6}$

The reduction formula is 
$$W=R(\xi)\exp{i[\alpha (\xi)+Bt+a \ln x ]}, \, \, \xi={y \over x}, \, \, \eta =\arctan {y \over x}=\phi \eqn\ewtwosix$$
and $\alpha$ satisfies
$$\alpha_{\xi}={S \over 1+\xi^2}{(1+R^2)^2 \over R^2}+{a\xi \over 1+\xi^{2}} \eqn\ealphadotsix$$
and $U(\phi)=-R^{2}(\xi)$ satisfies \eumultipl\ with $K_{1}=K_{2}=-{K \over 2}+4S^2+2a^2, K_{3}=-4S^{2}$.
For $a=0$ the equation again reduces to \erootone\ .

A real solution is obtained only for $K > 8S^2$, namely \eblafive\ . It is periodic in $\phi$ and hence singlevalued when $\mu$ is an integer.

For $S=0$ we get a real solution for $K_{1} < 0$, namely solution \eblablafour\  which in this case  reduces to
$$U(\phi)=-\tan^2{1 \over 2}\sqrt{|K_{1}|} (\phi -\phi_{0}). \eqn\eUsixSzero$$
This is a singlevalued function whenever $\sqrt{|K_{1}|}$ is an integer.

6. Algebras $A_{2,7}$ and $A_{2,8}$

The corresponding reductions lead to equations that we cannot decouple without introducing higher 
derivatives, so we will not treat them here.


\chapter{Solutions of the Nonlinear Diffusion Equation}

\section{General Procedure}

Let us consider the system \ephasediffusion\ and \eradialdiffusion\ , the NDLE for short. We impose that the solution be invariant under one of the Lie groups generated by the algebra in eq. \eeverything\ . The functions $R$ and $Q$ will then have the form \eRandQ\ with $\xi$ and $\beta$ as in Section 3 (different for each subalgebra $A_{2,1}, \ldots ,A_{2,8}$).

As for the LL equation, algebra $A_{2,1}$ must be treated separately.

For $A_{2,2}, \ldots , A_{2,8}$ we always have $(\nabla\xi)^2 \neq 0$. Eq. \ephasediffusion\ and \eradialdiffusion\ reduce to
$$\alpha_{\xi\xi}=-2{(1-R^2) \over 1+R^2} {R_{\xi} \over R} \alpha_{\xi} +
{f_{\xi} \over f} \alpha_{\xi} -2fm {(1-R^2) \over 1+R^2} {R_{\xi} \over R}
+hf \eqn\ealphanlde$$
$$R_{\xi\xi}={f_{\xi} \over f} R_{\xi} +{2R \over 1\!+\!R^2 } R_{\xi}^2 +{1\!-\!R^2 \over 1\!+\!R^2} R \bigl[\alpha_{\xi}^2 +2mf\alpha_{\xi} +g^2 \bigl] +{1 \over (\nabla\xi)^2} \bigl[BR+AR{1\!-\!R^2 \over 1\!+\!R^2} \bigr] \eqn\eRnlde$$

The functions $f(\xi), m(\xi), h(\xi)$ and $g(\xi)$ are defined by the relations
$$\eqalign
{ {f_{\xi} \over f}={\xi_{t} - \Delta\xi \over (\nabla\xi)^2}, \quad &m={(\nabla\xi,\nabla\beta) \over (\nabla\xi)^2f} \cr
h={\beta_{t}-\Delta\beta \over (\nabla\xi)^2f}, \quad &g^2={(\nabla\beta)^2 \over (\nabla\xi)^2} \cr}\eqn\efmhg$$

In order to decouple equations \ealphanlde\ and \eRnlde\ , we impose a restriction on the functions defined in  \efmhg\ namely
$$m_{\xi}+h =0 \eqn\erestrictionmh$$
Eq. \ealphanlde\ can then be integrated once to give 
$$\alpha_{\xi}=\bigl[S{(1+R^2)^2 \over R^2} -m \bigl] f(\xi)\eqn\ealphaintegrated$$
where $S$ is an integration constant. We substitute \ealphaintegrated\ into \eRnlde\ and put
$$R(\xi) =\sqrt{-U(\eta)}, \qquad \eta=\int f(\xi)d\xi. \eqn\eRtoUnlde$$

The equations are decoupled and the one for $U(\eta)$ is already in a standard form \qquad 
\Ref\rince{E.L.~Ince: Ordinary Differential Equations. New York: Dover 1956}  namely
$$U_{\eta\eta}=\bigl({1 \over 2U}\! +\!{1 \over U\!-\!1} \bigr)U_{\eta}^2 +2S^2{(1\!+\!U)(1\!-\!U)^3 \over U} +M{1\!+\!U \over 1\!-\!U } U\! +\!NU \eqn\eUetaeta$$
with
$$\eqalign
{M=&{2 \over f^2} \bigl(g^2 -m^2f^2+{A \over (\nabla\xi)^2} \bigr) \cr 
N=&{2 \over f^2} {B \over (\nabla\xi)^2}. \cr}
\eqn\eMN$$
We now make a further restriction, namely, that $M$ and $N$, defined in
 \eMN\ are constants. Eq. \eUetaeta\ then has a first integral $K$ and we obtain the elliptic equation \eudotpolynomial\ with
$$\eqalign
{K_{1}=&-{K \over 2} +4S^2 +M +N \cr
K_{2}=& -{K \over 2} +4S^2 +M -N \cr
K_{3}=& -4S^2. \cr}
\eqn\eabgnlde$$

In many cases we have $M=N=0$ and the polynomial on the right hand side of  
\eudotpolynomial\ has a double root at $U_{3}=U_4=1$. The solution we obtain for $S \neq 0$ is \eblafive\  with $U_4=1$ ie:
$$\eqalign
{U(\eta)=& 1-{(1-U_{2})(1-U_{1}) \over (U_{2}-U_{1})\sin^2\mu(\eta-\eta_{0})+1-U_{2}} \cr
\mu=& S \sqrt{(1-U_{2})(1-U_{1})},\quad U_{1}\leq U \leq U_{2} < 0 \cr}\eqn\eblafivendle$$
For $S=0$, $K>0$ we obtain solution \eblablafour\ i.e
$$U=- \tan ^2 {\sqrt{{K \over 8}}(\eta-\eta_{0})} \eqn\eSzeronlde$$

\section{Individual reductions} 
\item{1.} Algebra $A_{2,1}$ \hfil\break
We have 
$$R=R(t), \qquad Q=\alpha_{0} +ax+by \eqn\etwoonenlde$$
where $\alpha_{0}$ is a constant and $R_{t}$ satisfies:
$$R_{t}= -{R(1-R^2) \over 1+R^2} (A+a^2+b^2) -BR \eqn\etwoonenldea$$
Equation \etwoonenldea\ can easily be integrated (differently depending on whether $(A+B+a^2+b^2)(A-B+a^2+b^2)$ vanishes, or not) and we obtain a transcendental equation for $R(t)$.

\item{2.} Algebra $A_{2,2}$ \hfil\break

The reduction formula is \eatwotwo\ . We have $m=0$ and \erestrictionmh\
requires $b=0$ so the solutions are static ones. The variable $\eta$ and constants involved satisfy
$$\matrix
{&\eta=\ln\rho , \quad M=2a^2 \cr
&A=B=N=b=0 \cr}
\eqn\etwotwonlde$$
Since we have $b=0$ , the solutions are static ones. All solutions \eblaone\ , $\ldots$ , \eblablanine\ can occur.

\item{3.} Algebra $A_{2,3}$ \hfil\break
We have 
$$\eta=x, \quad M=2(b^2+A), \quad N=2B, \quad a=0 \eqn\etwothreenlde$$
and again, the solutions are static ones, since the reduction formula is \ewtwofour. All solutions of Section 3.2 can occur.

\item{4.} Algebra $A_{2,4}$ \hfil\break
The reduction formula is \ewtwofour\ and we have
$$\eqalign{
\eta=&{1 \over v} (e^{v(x+vt)} -1), \quad a=0, \quad A=-b^2, \quad B=0 \cr
M=&N=0, \quad v \neq 0\cr}
\eqn\etwofournlde$$
The obtained solutions are \eblafivendle\ and \eSzeronlde\
\quad and they are $t$-dependent.

\item{5.} Algebra $A_{2,5}$ \hfil\break
We have eq. \ewtwofive\ with
$$\eqalign
{\eta&=\xi=\,\ln\rho+{1 \over b} \phi , \quad A=B=N=0, \cr
M&={2a^2b^2 \over (b^2+1)^2}, \quad b  \neq  0 \cr} \eqn\etwofivenlde$$
Since we have $M \geq 0$ we obtain the solutions \eblafive, \eblaseven, \eblanine, \eblablaone, \eblablatwo, \eblablafive, \eblablasix\ and
\eblablanine.
The parameters of these solutions must satisfy, however, certain constraints
in order for the solutions to be single-valued. For $K=2M$ we obtain
$$U=-R_{0}^2 \rho^{{2|ab| \over b^2+1}} e^{{2|a| \over b^2+1} \phi} \eqn\etwofivenlde$$
a solution that is not singlevalued.

\item{6.} Algebra $A_{2,6}$ \hfil\break
The reduction formula is \ewtwosix\ and we have
$$\eta=\int {d\xi \over 1+\xi^2} =\phi, \quad A=B=a=M=N=0 \eqn\etwosixnlde$$
so the relevant solutions are \eblafivendle\ and \eSzeronlde\
The solutions are static and they are singlevalued for $\mu$ or $\sqrt{K/8}$ being integers.

\item{7.} Algebra $A_{2,7}$ \hfil\break
We put
$$\eqalign
{W=&R(\xi)\exp{i[\alpha(\xi)+{a \over 2}\ln t +b\phi]}\cr
\xi=&{x^2+y^2 \over t}, \quad \eta=\int {1 \over \xi} e^{-{\xi \over 4}} d\xi = Ei(-{1 \over 4}\xi) \cr}
\eqn\etwosevenndle$$
where $Ei(x)$ is the exponential integral function.
Moreover, we have $b=a=A=B=M=N=0$ and the relevant solutions are \eblafivendle\ and \eSzeronlde\

\item{8.} Algebra $A_{2,8}$ \hfil\break
We have
$$\eqalign{
W=&R(\xi)\exp{i[\alpha(\xi)+{a \over 2}\ln t]}, \quad \xi={x \over \sqrt{t}} \cr
\eta=&\int e^{-\xi^2/4} d\xi=\sqrt{\pi}\Phi(\xi) \cr }
\eqn\etwoeightndle$$
where $\Phi(x)$ is the probability integral. We have $M=N=A=B=a=0$ and 
so we obtain solutions \eblafivendle\ and \eSzeronlde\

We see that time-dependent solutions are obtained for the algebras $A_{2,1}, A_{2,4}, \break A_{2,7}$ and $A_{2,8}$. For $A_{2,5}, A_{2,7}$ and $A_{2,8}$ the solutions are trigonometric ones.

The phases $\alpha(\xi)$ can be calculated by direct integration, since we have
$${d\alpha \over d\eta}=-S{(1-U)^2 \over U} \eqn\etwoeightndlea$$
and $U$ is already known. 

Thus for $U$ given by \eblafivendle\ we get
$$\eqalign{\alpha(\eta)&={\sqrt{(U_1-1)(U_2-1)\over U_1U_2}}\;\arctan\Bigr\{\sqrt{U_2(U_1-1)\over U_1(U_2-1)}\cot{\mu(\eta-\eta_{0})}\Bigr\}\cr
&-{\sqrt{(U_1-1)(U_2-1)}}\;\arctan\Bigr\{\sqrt{(U_1-1)\over
(U_2-1)}\cot{\mu(\eta-\eta_{0})}\Bigr\}\,+\,\alpha_0,
\cr}\eqn\eephaseone$$
while for $U$ given by
 \eSzeronlde\ we have
$$\alpha=4\sqrt{2\over K}S\,\cot\{\sqrt{2K}(\eta-\eta_0)\}\,+\,\alpha_0.
\eqn\ephaseb$$


\chapter{Conclusions}
The Landau-Lifshitz equation \eun has received quite a bit of previous attention, 
mainly in the context of continuum Heisenberg ferromagnetic spin systems
\Ref\rlak{M.~Lakshmanan: Phys.Lett. A 61  (1977) 53}
\Ref\rzaktak{V.E.~Zakharov, L.A.~Takhtadzhyan: Theor. Mat. Fiz  38 (1979) 26}
\Ref\rlakdan{M.~Lakshmanan, M.~Daniel: Physica A 107 (1981) 533}
\Ref\rmikhyar{A.V.~Mikhailov, A.I.~Yaremchuk: JETP Lett. 36 (1982) 78}
\Ref\rlakpor{M.~Lakshmanan, K.~Porsezian: Phys. Lett. A 146 (1990) 329}
\Ref\rporlak{K.~Porsezian, M.~Lakshmanan: J. Math. Phys. 32 (1991) 2923}
\Ref\rdanpor{M.~Daniel, K.~Porsezian, M.~Lakshmanan: J. Math. Phys. 35 (1994) 6498}.
 The anisotropy coefficients A and B of eq. \etrois were usually set equal to zero. Use was made 
 of the fact that eq \eun is integrable, at least in the one-dimensional case, or in the
 two-dimensional, spherically symmetric one.
 
 A one-soliton solution has been obtained \refmark\rlakdan  \refmark\rlakpor a radially symmetric one. In our variables $R, Q$ of eq. \eradiallandau , \ephaselandau 
 this solution corresponds to
 $$R={ 4t+\alpha_{1} \over 4t+\alpha_{1}+\cosh^{2}\bigl[{4t+\alpha_{1} \over (4t+\alpha_{1})^{2}+\alpha_{2}^{2}
 }(x^{2}+y^{2}) \bigr]} \eqn\econclone$$
 where $\alpha_{1}$ and $\alpha_{2}$ are arbitrary real constants. As noted by Lakshamanan and Porsezian
 \refmark\rlakpor the soliton spreads in time.
 
 The solution \econclone\ is not among the invariant solutions obtained in this article, nor can it be 
 obtained from such a solution by applying transformations from the symmetry group. As often happens
\Ref\rlmw{D.~Levi, C.R.~Menyuk, P.~Winternitz: Phys. Rev. A 44 (1991) 6057},
the method of symmetry reduction that does not rely on integrability, provides different solutions for 
integrable equations, than the use of Lax pairs, or Backlund transformations.

We  note that eq. \eun\ with $A \neq 0$ is not integrable.

In general, we have reduced the LL equation to the ordinary differential equation \eudotdot. 
We have integrated eq. \eudotdot\ in terms of elliptic functions whenever $p,q$ and $m$ are 
constants. For the algebras $A_{2,3}, \ldots , A_{2,6}$ this was always the case.

For algebra $A_{2,2}$ we obtained eq. \eudotpolynomial\ only for $A=0, B=b$. Let us briefly consider
the case when the anisotropy coefficient $A$ does not vanish. We then return to the 
 original variable $\xi=\rho=\sqrt{x^{2}+y^{2}}=exp \eta $ and transform eq. \eudotdot\ into
 $$U_{\rho\rho}=({1 \over 2U}+{1 \over U-1})U_{\rho}^2-{1 \over \rho}U_{\rho}+{2S^{2} \over \rho^{2}}
 (U-1)^2(-U+{1 \over U}) -{2(a^{2}+A^{2}\rho^{2}) \over \rho^{2}}+2(B-b) \eqn\econcltwo$$
 For $a=0,b=B$ this is the equation for the fifth Painlev{\'e} transcendent $P_{V}$ [\rince]. However,
 for $B \neq b$ eq \econcltwo\ does not have the Painlev{\'e} property. According to the Painlev{\'e}
  conjecture 
\Ref\rars{M.J.~Ablowitz, A.~Ramani and H.~Segur: J. Math. Phys 21 (1980) 1006}
\Ref\rabclark{M.J.~Ablowitz, P.A.~Clarkson: Solitons, Nonlinear Evolution Equations and 
Inverse Scattering. Cambridge: Cambridge University Press 1991}, eq. \eun\ is hence,
in general, not integrable.

This has not stopped us from obtaining numerous solutions, both in integrable and nonintegrable
cases. The algebra $A_{2,2}$ (cylindrical symmetry) for $A=0, B=b$ leads to periodic 
solutions, as discussed in Section 3.3. The periodicity is in the radial variable $\rho$.
The time dependence is restricted to the phase $Q$, as is seen in eq. \eatwotwo . Moreover
the time-dependence is entirely due to the presence of the external field $B$
(we have $b=B$) that generates a rotation between the components $\phi_{1}$ and $\phi_{2}$
of the original vector $\vphi$.

Some elementary nonperiodic solutions that we can extract from Section 3 are

$$R^{2}={U_{1}-U_{2}S^{2}(U_{2}-U_{1}(ln \rho /\rho_{0})^{2} \over 1+S^{2}(U_{2}-U_{1})^{2}(ln\rho/
\rho_{0})^{2}} \eqn\econclthree$$
$$R^{2}={4R_{0}^{2}U_{1}(U_{2}-U_{1})-U_{2}(U_{4}-U_{1})[2R_{0}^{2}+\rho^{2\mu}+R_{0}^{4}\rho^{-2\mu} \over (U_{4}-U_{1})[2R_{0}^{2}+\rho^{2\mu}+R_{0}^{4}\rho^{-2\mu}]-4(U_{2}-U_{1})R_{0}^{2}} \eqn\econclfour$$
with $S,R_{0},\rho_{0}, U_{i}$ constants and
$$Q=S \int {{(1+R^{2})^{2}} \over {\rho R^{2}}} d\rho +a\phi +Bt \eqn\econclfive$$
in both cases.

For $S=0$ we have for instance
$$R=[-U_{1}+ \sqrt{2 \over -K_{1}} ({ln {\rho \over \rho_{0}})^{-1}}]^{1/2} \eqn\econclsix$$
$$R={2 \sqrt{-U_{1}} R_{0} \over {\rho^{\mu}+R_{0}^{2}\rho^{-\mu}}} \eqn\econclseven$$
with $$Q=a\phi+Bt+Q_{0} \eqn\econcleight$$.

For $A \neq 0$, as mentioned above , solutions are obtained in terms of $P_{V}(\rho)$. Their 
time dependence is again given  by the term $Bt$ in the phase $Q$.

For algebras $A_{2,3}$ and $A_{2,4}$ we obtain eq. \eudotpolynomial and a multitude of explicit 
solutions for all values of $a,b,A$ and $B$ . Note that for
$$S \neq 0, \quad B=a, \quad A=-b^{2} \eqn\econclnine$$
in particular for the one dimensional $(b=0)$, static $(a=0)$ with no external fields $(A=B=0)$, two
 of the roots in eq. \eumultipl coincide and the equation reduces to
 $$U_{\eta}^{2}=-4S^{2}(U-1)^{2}[U^{2}+(-{K_{1} \over 4S^{2}} +2)U-{K_{3} \over 4S^{2}}] \eqn\econclten$$
 
Eq. \econclten\ only allows elementary solutions like (3.12), ...(3.17), not however
the elliptic function ones. These occur when the fields $A$ and $B$ are such that \econclnine\ is
not satisfied.

To our knowledge, the NLDE \edeux\ has not been investigated from the point of view of its
integrability and we have no solutions to compare ours to.

We have derived many explicit exact solutions of both equations. 
Looking at them we note that most of them have infinite
energy. They can describe coherent phenomena in various
solid state and condensed matter applications.

Looking first
at the solutions of the LL equation we note that
some of our solutions have finite energy. In particular, this is the case
for  \efourtyone. This solution is obtained from the familiar
static solution describing $n$ solitons ``on top of each other"\Ref\rwoj{see \eg\
B.~Piette, W.J.~Zakrzewski: Skyrmion Dynamics in (2+1) Dimensions 
to appear in  Chaos, Solitons
and Fractals (1995)}. Its time dependence is given by
the factor $e\sp{iBt}$ which thus describes a rotation of this static
solution in the $\phi_1$, $\phi_2$ plane with the angular frequency 
given by the anisotropy $B$. The other solutions of this class
correspond to the static elliptic solutions discussed in ref [\rourpaper]
again rotated by $e\sp{iBt}$.

The solutions corresponding to algebras $A_{2,1}$, $A_{2,3}$
and $A_{2,4}$ have infinite energies. As such, they describe
various waves in the medium (generalisations of plane waves).
These can for instance be spin waves; the energy per period is finite.

An interesting class of solutions are those corresponding to algebras
$A_{2,5}$ and $A_{2,6}$. Given the choice of parameters, these
solutions can be of finite energy; however, due to their dependence
on the variable $\phi=\arctan{y\over x}$, they may become singular
when $x$ and $y$ vanish. They can be used to describe
media with defects.

Most of the comments made above apply also to the solutions 
of the NLDE. The static solutions in both cases are of course the same.
When we consider nonstatic solutions, the most interesting  from the
physical point of view, are perhaps solutions corresponding
to algebras $A_{2,1}$, $A_{2,7}$ and $A_{2,8}$. All of them
have infinite energies. The solution corresponding to $A_{2,1}$
describes a structure shrinking towards the origin
(or expanding to infinity - depending on the values of the parameters).
The other solutions describe field
configurations evolving in time. They can be used in the description of some
physical phenomena in condensed matter or solid state physics.

Among the questions that we plan to return to, we mention the study of "partially
invariant" solutions
\Ref\rovs{L.V.~Ovsiannikov: Group Analysis of Differential Equations. New York:
Academic Press 1982}
\Ref\martw{L.~Martina, P.~Winternitz: J. Math. Phys. 33 (1992) 2718}
\Ref\rmartsw{L.~Martina, G.~Soliani, P.~Winternitz: J. Phys. A 25 (1992)
4425}
 of eq. \eun and \edeux , and also "conditionally invariant" ones
\Ref\rlevidw{D.~Levi, P.~Winternitz: J.Phys A 22 (1989) 2915}.
A study of solutions involving Painlev{\'e} transcendents is also warranted.

\ack
\par
Most of the work reported in this paper was performed when one of us (WJZ)
visited the CRM, Universit\'e de Montr\'eal, Canada.
He wishes to thank  the Centre de Recherches Math\'ematiques for its support and hospitality.
The research of P.W. was partly supported by research grants from NSERC of Canada
and FCAR du Qu{\'e}bec.
\vfil
\refout

\end
\bye